\documentclass{article}

\usepackage{color}
\usepackage{amsmath,amsthm}
\usepackage{amsfonts,amssymb}
\usepackage{stmaryrd,enumitem}

\definecolor{myurlcolor}{rgb}{0.6,0,0}
\definecolor{mycitecolor}{rgb}{0,0,0.8}
\definecolor{myrefcolor}{rgb}{0,0,0.8}
\usepackage[bookmarks=false,pagebackref=true]{hyperref}
\hypersetup{colorlinks,
linkcolor=myrefcolor,
citecolor=mycitecolor,
urlcolor=myurlcolor}

\newcommand{\define}[1]{{\bf \boldmath{#1}}}
\newcommand{\maps}{\colon}

\newcommand{\tr}{\mathrm{tr}}
\newcommand{\Aut}{\mathrm{Aut}}

\newcommand{\R}{\mathbb{R}}
\newcommand{\C}{\mathbb{C}}
\renewcommand{\H}{\mathbb{H}}
\renewcommand{\O}{\mathbb{O}}
\newcommand{\K}{\mathbb{K}}
\newcommand{\h}{\mathfrak{h}}

\theoremstyle{plain}

\newtheorem{thm}{Theorem}

\newtheorem{ex}[thm]{Example}

\theoremstyle{definition}

\newcommand{\beq}{\begin{equation}}
\newcommand{\eeq}{\end{equation}}

\usepackage{tikz}
\usetikzlibrary{backgrounds,circuits,circuits.ee.IEC,shapes,fit,matrix}
\usetikzlibrary{arrows,positioning,fit,matrix,shapes.geometric,external}

\tikzstyle{none}=[inner sep=0pt]
\tikzstyle{inarrow}=[->, >=stealth, shorten >=.2cm,line width=1.5]
\begin{document}

\begin{center}   
  {\bf Getting to the Bottom of Noether's Theorem \\}   
  \vspace{0.3cm}
  {\em John\ C.\ Baez \\}
  \vspace{0.3cm}
  {\small
 Department of Mathematics \\
    University of California \\
  Riverside CA, USA 92521 \\}
  \vspace{0.3cm}   
  {\small email:  baez@math.ucr.edu \\} 
  \vspace{0.3cm}   
  {\small Februrary 15, 2022}
  \vspace{0.3cm}   
\end{center}   

\begin{abstract}
We examine the assumptions behind Noether's theorem connecting symmetries and conservation laws.  To compare classical and quantum versions of this theorem, we take an algebraic approach.  In both classical and quantum mechanics, observables are naturally elements of a Jordan algebra, while generators of one-parameter groups of transformations are naturally elements of a Lie algebra.    Noether's theorem holds whenever we can map observables to generators in such a way that each observable generates a one-parameter group that preserves itself.  In ordinary complex quantum mechanics this mapping is multiplication by $\sqrt{-1}$.   In the more general framework of unital JB-algebras, Alfsen and Shultz call such a mapping a `dynamical correspondence', and show its presence allows us to identify the unital JB-algebra with the self-adjoint part of a complex C*-algebra.   However, to prove their result, they impose a second, more obscure, condition on the dynamical correspondence.  We show this expresses a relation between quantum and statistical mechanics, closely connected to the principle that `inverse temperature is imaginary time'.
\end{abstract} 

\vskip 1em

It is sometimes said Noether showed \emph{symmetries give conservation laws}.   But this is only true under some assumptions: for example, that the equations of motion come from a Lagrangian.  For which types of physical theories do symmetries give conservation laws?  What are we assuming about the world, if we assume it is described by a theories of this type?  It is hard to get to the bottom of these questions, but it is worth trying.

We can prove versions of Noether's theorem relating symmetries to conserved quantities in many frameworks.  While a differential geometric framework is truer to Noether's original vision, I will study this subject \emph{algebraically}, using ideas from Hamiltonian mechanics.  

Atiyah \cite{Atiyah} warned us of the dangers of algebra:
\begin{quote}
`` ...algebra is to the geometer what you might call the Faustian
offer. As you know, Faust in Goethe's story was offered whatever he wanted
(in his case the love of a beautiful woman), by the devil, in return for selling
his soul. Algebra is the offer made by the devil to the mathematician. The
devil says: I will give you this powerful machine, it will answer any question
you like. All you need to do is give me your soul: give up geometry and
you will have this marvellous machine.''
\end{quote}
While this is sometimes true, algebra is more than a computational tool: it lets us express concepts in a very clear and distilled way.  Further, the geometrical framework developed for classical mechanics is not sufficient for quantum mechanics.   An algebraic approach emphasizes the similarity between classical and quantum mechanics, clarifying their differences.       

Our algebraic study of Noether's theorem relies on an interlocking trio of important concepts used to describe physical systems: `states', `observables' and `generators'.    A physical system has a convex set of states.    An observable is a real-valued quantity whose value depends---perhaps with some randomness---on the state.  More precisely, an observable maps each state to a probability measure on the real line.  A generator, on the other hand, is something that gives rise to a one-parameter group of transformations of the set of states---or dually, of the set of observables.  

For example, when we say ``the energy of the system is 7 joules'', we are treating energy as an observable.   When we say ``the Hamiltonian generates time translations'', we are treating the Hamiltonian as a generator.    In both classical mechanics and ordinary complex quantum mechanics we usually say the Hamiltonian \emph{is} the energy, because we have a way to identify them.   But it has repeatedly been noted that observables and generators play distinct roles \cite{GrginPetersen, Landsmaan, Zalamea}---and in some theories, such as real or quaternionic quantum mechanics, they are truly different.   In all the theories considered here the set of observables is a Jordan algebra, while the set of generators is a Lie algebra. 

When we can identify observables with generators, we can state Noether's theorem as the following equivalence:
\begin{center}
The generator $a$ generates transformations that leave the
observable $b$ fixed.
\end{center}
\[ \Updownarrow \]
\begin{center} The generator $b$ generates 
transformations that leave the observable $a$ fixed.
\end{center}
In this beautifully symmetrical statement, we switch from thinking of $a$ as the generator and $b$ as the observable in the first part to thinking of $b$ as the generator and $a$ as the observable in the second part.   Of course, this statement is \emph{true} only under some conditions, which 
we explore in detail.   But the most fundamental condition, we claim, is the ability to identify
observables with generators.

In Section \ref{sec:classical} we explain how observables and generators are unified in classical mechanics. Here we treat observables as being \emph{the same} as generators, by treating them as elements of a Poisson algebra, which is both a Jordan algebra and a Lie algebra.

In Section \ref{sec:quantum} we turn to quantum mechanics.  Here observables are not quite the same as generators.  They are both elements of a $\ast$-algebra: observables are self-adjoint, obeying $a^* = a$, while generators are skew-adjoint, obeying $a^* = -a$.   The self-adjoint elements form a Jordan algebra, while the skew-adjoint elements form a Lie algebra.   

In ordinary complex quantum mechanics we use a \emph{complex} $\ast$-algebra.   This lets us turn any self-adjoint element into a skew-adjoint one by multiplying it by $\sqrt{-1}$.    Thus, \emph{the complex numbers let us identify observables with generators}.  In real and quaternionic quantum mechanics this identification is impossible.  Thus, the appearance of complex numbers in quantum mechanics is closely connected to Noether's theorem.

In short, the picture painted by Sections \ref{sec:classical} and \ref{sec:quantum} is this:

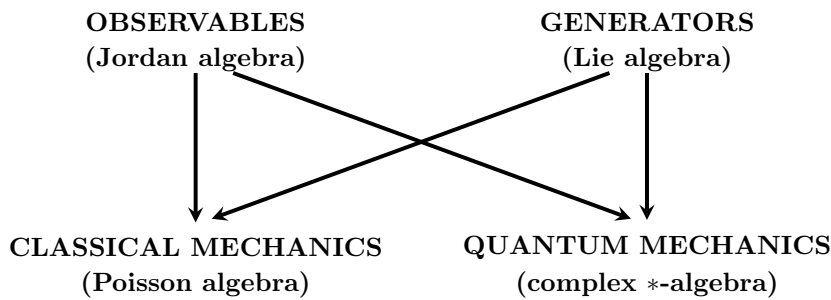
\[
\begin{tikzpicture}
	
		\node [style=none] (O) at (-3, 0) {\textbf{OBSERVABLES}};
		\node [style = none] (J) at (-3, -0.5) {\textbf{(Jordan algebra)}};
		\node [style=none] (G) at (3, 0) {\textbf{GENERATORS}};
		\node [style = none] (L) at (3, -0.5) {\textbf{(Lie algebra)}};
		
		\node [style=none] (C') at (-3, -2.7) {};
		\node [style=none] (C) at (-3, -3) {\textbf{CLASSICAL MECHANICS}};
		\node [style=none] (P) at (-3, -3.5) {\textbf{(Poisson algebra)}};
		\node [style=none] (Q') at (3, -2.7) {};
		\node [style=none] (Q) at (3, -3) {\textbf{QUANTUM MECHANICS}};
		\node [style=none] (S) at (3, -3.5) {\textbf{(complex $\ast$-algebra)}};
	
		\draw [style=inarrow] (J) to (C);
		\draw [style=inarrow] (J) to (Q'); 
		\draw [style=inarrow] (L) to (C');
		\draw [style=inarrow] (L) to (Q);
\end{tikzpicture}
\]

In Section \ref{sec:lie} we examine generators on their own: that is, Lie algebras.   Lie algebras arise very naturally from the concept of `symmetry'.  Any Lie group gives rise to a Lie
algebra, and any element of this Lie algebra then generates a one-parameter family of transformations of that very same Lie algebra.   This lets us state a version of Noether's theorem solely in terms of generators:
\begin{center}
The generator $a$ generates transformations that leave the generator $b$ fixed.
\end{center}
\[  \Updownarrow \]
\begin{center} The generator $b$ generates 
transformations that leave the generator $a$ fixed.
\end{center}
Furthermore, when we translate these statements into equations, their equivalence follows directly from this elementary property of the Lie bracket:
\[    [a,b] = 0 \iff [b,a] = 0.\]
Thus, Noether's theorem is almost automatic \emph{if} we forget about observables and work solely with generators.  The only questions left are: why should symmetries be described by Lie groups, and what is the meaning of this property of the Lie bracket?  In Section \ref{sec:lie} we say a bit about both these questions.   In particular, the Lie algebra formulation of Noether's theorem follows from a more primitive group formulation, which says that for any two group elements $g$ and $h$, 
\begin{center}
$g$ commutes with $h$ $\iff$ $h$ commutes with $g$.
\end{center}
That is: given two ways of transforming a physical system, the first is invariant under the second if and only if the second is invariant under the first.

However, observables are crucial in physics.   Working solely with generators in order to make Noether's theorem a tautology would be another sort of Faustian bargain.  To really get to the bottom of Noether's theorem, we need to understand the map from observables to generators.  In ordinary quantum mechanics this comes from multiplication by $i$.  But this just pushes the mystery back a notch: why should we be using the complex numbers in quantum mechanics?

In Section \ref{sec:jordan} we tackle this issue by examining observables on their own: that is, Jordan algebras.   Those of greatest importance in physics are the `unital JB-algebras', unfortunately named not after the author, but Jordan and Banach.   These allow a unified approach to real, complex and quaternionic quantum mechanics, along with some more exotic theories.  Thus, they let us study how the role of complex numbers in quantum mechanics is connected to Noether's theorem.

Any unital JB-algebra $O$ has a partial ordering: that is, we can talk about one observable being greater than or equal to another.  With the help of this we can define states on $O$, and prove that any observable maps each state to a probability measure on the real line.   More surprisingly, any JB-algebra also gives rise to \emph{two} Lie algebras.  The smaller of these, say $L$, has elements that generate transformations of $O$ that preserve all the structure of this unital JB-algebra.   They also act on the set of states.  Thus, elements of $L$ truly deserve to be considered `generators'.   

In a unital JB-algebra there is not always a way to reinterpret observables as generators. However, Alfsen and Shultz \cite{AlfsenShultz98,AlfsenShultz03} have defined the notion of a `dynamical correspondence' for such an algebra, which is a well-behaved map $\psi \maps O \to L$. 
One of the two conditions they impose on this map implies a version of Noether's theorem. They prove that any JB-algebra with a dynamical correspondence gives a complex $\ast$-algebra where the observables are self-adjoint elements, the generators are skew-adjoint, and we can convert observables into generators by multiplying them by $i$.  

This result is important, because the definition of JB-algebra does not involve the complex numbers, nor does the concept of dynamical correspondence.  Rather, the role of the complex numbers in quantum mechanics \emph{emerges} from a map from observables to generators that obeys conditions including Noether's theorem! 

To be a bit more precise, Alfsen and Shultz's first condition on the map $\psi \maps O \to L$ says
that every observable $a \in O$ generates transformations that leave $a$ itself fixed.  We call this the `self-conservation principle'.   It implies Noether's theorem.

However, in their definition of dynamical correspondence, Alfsen and Shultz also impose a second, more mysterious condition on the map $\psi$.   In Section \ref{sec:statistical} we argue that this condition is best understood in terms of the larger Lie algebra associated to a unital JB-algebra.  This is the direct sum $A = O \oplus L$, equipped with a Lie bracket such that 
\[    [-,-] \maps L \times L \to L    \qquad [-,-] \maps L \times O \to O \]
\[   \,\, [-,-] \maps O \times L \to O    \qquad [-,-] \maps O \times O \to L .\]
As mentioned, elements of $L$ generate transformations of $O$ that preserve all the structure
on this unital JB-algebra.   Elements of $O$ also generate transformations of $O$, but these only preserve its vector space structure and partial ordering.

In Section \ref{sec:statistical} we argue that these other transformations are connected to \emph{statistical} mechanics.  For example, consider ordinary quantum mechanics and
let $O$ be the unital JB-algebra of all bounded self-adjoint operators on a complex Hilbert
space.  Then $L$ is the Lie algebra of all bounded skew-adjoint operators on this Hilbert space. There is a dynamical correpondence sending any observable $H \in O$ to the generator $\psi(H) = iH \in L$, which then generates a one-parameter group of transformations of $O$ defined as follows:
\[             a \mapsto e^{itH} a e^{-itH}  \qquad \forall t \in \R, a \in O.  \]
If $H$ is the Hamiltonian of some system, this is the usual formula for time evolution of observables in the Heisenberg picture.   But $H$ also generates a 
one-parameter group of transformations of $O$ as follows:
\[           a \mapsto  e^{-\beta H/2} a e^{-\beta H/2}  \qquad \forall \beta \in \R, a \in O .\] 
Writing $\beta = 1/kT$ where $T$ is temperature and $k$ is Boltzmann's constant, we argue that these are `thermal transformations'.    Acting on a state in thermal equilibrium at some temperature, these transformations produce states in thermal equilibrium at other temperatures (up to normalization).  

The analogy between $it$ and $1/kT$ is often summarized by saying `inverse temperature is imaginary time'.   The second condition in Alfsen and Shultz's definition is a way of capturing this principle in a way that does not explicitly mention the complex numbers.  Thus, we may \emph{very roughly} say their result explains the role of complex numbers in quantum mechanics starting from three assumptions:
\begin{itemize}
\item observables form a unital JB-algebra,
\item the self-conservation principle (and thus Noether's theorem),
\item the principle that inverse temperature is imaginary time.
\end{itemize}

\subsubsection*{Warning}

The concept of `generating a one-parameter family of transformations' involves a differential equation.  This necessarily brings some analysis into our otherwise algebraic discussion.  Given a function $f \maps \R \to V$ from the real numbers to a vector space, the meaning of the derivative 
\[         \frac{df}{dt} = \lim_{h \to 0} \frac{f(t+h) - f(t)}{h}  \]
depends on the topology we place on $V$.   In some of our theorems we do not say what topology we are using.  Any vector space $V$ has a locally convex topology where a net $v_\alpha \in V$ converges to $v \in V$ if and only if $\ell(v_\alpha) \to \ell(v)$ for all linear functionals $\ell \maps V \to \R$, so we can use this, but in practice we often want to use another.    To focus on key ideas, we downplay this issue.  In Theorems \ref{thm:noether_1}, \ref{thm:noether_2} and \ref{thm:noether_3} we can use any locally convex topology such that the relevant bracket operation is continuous in each argument.   In Theorem \ref{thm:poisson} we should use the $C^\infty$ topology.   In Theorems \ref{thm:C*-algebra} and \ref{thm:noether_4} we should use the norm topology.

All the theorems in this paper are either easy or already known.  Our goal here is to deploy this mathematics to better understand Noether's theorem.

\subsubsection*{Acknowledgements}

I thank Javier Prieto for suggesting the connection to the work of Leifer and Spekkens, and Emilio Cobanera for spotting errors.

\section{Classical mechanics}
\label{sec:classical}

We begin with a lightning review of classical mechanics.  Noether's theorem is often studied in the Lagrangian approach to classical mechanics. Here we take the Hamiltonian approach, and refer the reader elsewhere for the link back to Lagrangian mechanics \cite{MarsdenRatiu}.
Our focus is on the dual role that observables play in Hamiltonian mechanics.  First, and most obviously, they assign a number to each state.  But second, they generate one-parameter groups.   This dual role underlies the Hamiltonian formulation of Noether's theorem, Theorem \ref{thm:noether_1}.

In the classical mechanics of a point particle in $\R^n$, the space of states is 
$\R^{2n}$, with coordinates $q_1, \dots, q_n$ representing the particle's
position and coordinates $p_1, \dots, p_n$ representing its momentum.   Observables---real-valued quantities depending on the particle's state---are functions on $\R^{2n}$.   
In order to differentiate these as many times as we want without running into trouble 
we take our observables to lie in $C^\infty(\R^{2n})$, the set of smooth real-valued functions on 
$\R^{2n}$.

The set $C^\infty(\R^{2n})$ is a real vector space where we can add and multiply functions
pointwise, and the laws of a commutative algebra hold:
\[     f(gh) = (fg)h  \qquad fg = gf \]
\[      f(\beta g + \gamma h) = \beta fg + \gamma fh  \]
for all $f,g,h \in C^\infty(\R^{2n})$ and $\beta, \gamma \in \R$.   The multiplication 
has a clear physical meaning: for example, to measure the observable $fg$ in some
state $x \in \R^{2n}$ is the same as to measure $f$ and $g$ separately in this state and then multiply the results.  Of course this becomes problematic in quantum mechanics, but here we assume that both observables can be measured simultaneously with perfect accuracy. 

To do interesting physics with our observables, and to state Noether's theorem, we need an additional operation on $C^\infty(\R^{2n})$.  This is the \define{Poisson bracket}, defined by
\[     \{f, g\} = \sum_{i = 1}^n \frac{\partial f}{\partial p_i}\frac{\partial g}{\partial q_i} 
- \frac{\partial f}{\partial q_i} \frac{\partial g}{\partial p_i} .\]
This gives a way for each observable to define a one-parameter family of transformations of all the rest, at least if a technical condition holds.   Most notably, when $f \in C^\infty(\R^{2n})$ is `energy', these transformations are `time evolution'.  

How does this work?    We say $f \in C^\infty(\R^{2n})$ \define{generates a one-parameter 
group} it there exist maps 
\[     F^f_t \maps C^\infty(\R^{2n}) \to C^\infty(\R^{2n})   \quad \forall t \in \R \]
obeying \define{Hamilton's equations}:
\[     \frac{d}{dt} F^f_t(g) =  \{f, F^f_t(g)\}   \qquad \forall t \in \R   \]
and the initial conditions
\[      F^f_0(g) = g \]
for all $g \in C^\infty(\R^{2n})$.   Such a family of maps may not exist, but conditions
are known under which it does: for example, when the partial derivatives of $f$ are bounded.  
When such a family exists, it is unique, and it obeys the equations
\[        \begin{array}{ccl}           F^t_0 &=& \mathrm{id}   \\ 
              F^f_{s+t} &=&  F^f_s \circ F^f_t  \qquad \forall s, t \in \R   \end{array} \]
which are the usual definition of a one-parameter group.   It follows from these equations
that each transformation $F^f_t$ is invertible.

\begin{ex}
Take $\R^2$ to be the space of states for a particle on a line, with coordinates
$(q,p)$.   Let $H \in \C^\infty(\R^2)$ be the energy of a particle of mass $m > 0$ in 
a smooth potential $V \maps \R \to \R$, namely
\[      H(q,p) = \frac{p^2}{2m} + V(q) .\]
The position and momentum of the particle at time $t$, if they are defined, are observables
given by 
\[     q_t = F^H_t(q), \qquad p_t = F^H_t(p)  .\]
Hamilton's equations say
\[    \frac{dq_t}{dt} = \{H, q_t\}  , \qquad \frac{dp_t}{dt} = \{H, p_t\} . \]
Working out the Poisson brackets, these give
\[    \frac{dq_t}{dt} = \frac{p_t}{m} , \qquad \frac{dp_t}{dt}= -V' \circ q_t  \]
as one would expect.   However, these equations may not have global-in-time solutions:
for example, when $V(q) = -q^2$ the particle will typically run off to infinity in a finite
amount of time.   But if $V$ is well-behaved---for example, when it has bounded derivative or 
is bounded below---Hamilton's equations have global-in-time solutions and $H$ generates a 
one-parameter group.
\end{ex}

Suppose that $f \in C^\infty(\R^{2n})$ generates a one-parameter group.  Then the 
transformations $F^f_t$ preserve all the structure on observables we have mentioned so far.
First, these transformations are linear.  One can prove this with a well-known calculation
using the linearity of the Poisson bracket in its second argument:
\[        \{f, \beta g + \gamma h\} = \beta \{f, g\} + \gamma \{f, h\}  .\]
Second, these transformations preserve multiplication, which one can prove using the \define{Leibniz law}:
\[      \{f, gh \} = \{f, g\} h + g \{f, h\}.   \]
And third, they preserve the Poisson bracket, which one can prove using the \define{Jacobi
identity}:
\[       \{f, \{g, h\}\} = \{ \{f, g\}, h\} + \{g , \{f, h\} \} .\]
We can summarize all this by saying the transformations $F^f_t$ are Poisson algebra
automorphisms.  

Finally, we have Noether's theorem relating symmetries and conserved quantities.  
Suppose $f \in C^\infty(\R^{2n})$ generates a one-parameter group.  We
say that $f$ generates \define{symmetries} of $g \in C^\infty(\R^{2n})$ if
\[                F^f_t(g) = g   \qquad \forall t \in \R  .\]
In this situation we also say $g$ is \define{conserved} by the group generated by $f$.
Both are ways of saying that the one-parameter group generated by $f$ leaves $g$
unchanged.   

Noether's theorem says that if both $f$ and $g$ generate one-parameter groups, then
\begin{center}
$f$ generates symmetries of $g$ $\iff$ $g$ generates symmetries of $f$.
\end{center}
In other words, $g$ is conserved by the group generated by $f$ if and only if $g$
generates symmetries of $f$.  The proof, which we give more generally below, relies crucially on one extra law obeyed by the Poisson bracket, its \define{antisymmetry}:
\[                  \{f, g \} = -\{g, f\}.  \]
We shall see that $\{f,g\} = 0$ iff $f$ generates symmetries of $g$, while $\{g,f\} = 0$
iff $g$ generates symmetries of $f$.

Everything so far can be abstracted and generalized, giving an algebraic approach to classical mechanics where we assume observables form a `Poisson algebra'.   A \define{Poisson algebra} is a real vector space $A$ equipped with a multiplication making $A$ into a \define{commutative algebra}:
\[     a(bc) = (ab)c  \qquad ab = ba \]
\[      a(\beta b + \gamma c) = \beta ab + \gamma ac  \]
together with a bracket making $A$ into a \define{Lie algebra}:
\[          \{a, \{b,c\} \} = \{ \{a, b\}, c \} + \{b, \{a,c\} \}  \qquad \{a,b\} = -\{b,a\} \]
\[       \{ a, \beta b + \gamma c \} = \beta \{a,b\} + \gamma \{a,c\}   \]
and obeying the \define{Leibniz law}:
\[           \{ a, bc \} = \{a,b\} c + b \{a,c\}   \]
for all $a,b,c \in A$ and $\beta, \gamma \in \R$.  

A typical example of a Poisson algebra is the algebra of smooth functions on a symplectic manifold.  More generally, a \define{Poisson manifold} is a smooth manifold $M$ such that $C^\infty(M)$, made into a commutative algebra in the usual way, is also equipped with a bracket making it a Poisson algebra.  For an introduction to symplectic and Poisson manifolds in physics, see \cite{GuilleminSternberg,MarsdenRatiu,Vaisman}.

Suppose $A$ is a Poisson algebra and $a \in A$.  Let us say that $a$ \define{generates a one-parameter family} if for each $b \in A$ there exists a unique solution $F_t^a(b)$ to Hamilton's
equations
\[     \frac{d}{dt} F_t^a (b) =  \left\{a, F_t^a (b) \right\}   \qquad \forall t \in \R   \]
obeying the initial conditions
\[      F_0^a (b) = b . \]
Note that we do not demand the one-parameter group law
\[                   F_s^a (F_t^a (b)) = F_{s+t}^a (b)  \qquad \forall s, t \in \R  .\]
In well-behaved situations this follows.  For example, we have:

\begin{thm}
\label{thm:poisson}
Let $M$ be a compact Poisson manifold and $A = C^\infty(M)$ the resulting Poisson
algebra.   Then every element $a \in A$ generates a one-parameter family $F_t^a 
\maps A \to  A$.   Furthermore, $F_t^a$ is a one-parameter group
of linear transformations preserving the multiplication and the Poisson bracket.
\end{thm}

\begin{proof}
Given $a \in A$ the operation $\{a,-\}$ is a derivation of $A$ and thus corresponds to a
smooth vector field $v$ on $M$.   Any smooth vector field on a compact manifold is integrable, so there is for any $x \in M$ a unique solution of
\[     \frac{d}{dt} \phi_t(x) = v(\phi_t(x))   \qquad \forall t \in \R  \]
obeying the initial conditions
\[     \phi_0(x) = x  \qquad \forall x \in X .\]
This defines a `flow', that is, a smooth one-parameter family of diffeomorphisms $\phi_t \maps M \to M$ obeying $\phi_s \circ \phi_t = \phi_{s+t}$ for all $s,t \in \R$.  We can define $F_t^a$ by 
\[    F_t^a(b) = b \circ \phi_t  \]
and this is a one-parameter group of linear transformations preserving the multiplication
in $A$.   These transformations also preserve the Poisson bracket \cite[Prop. 10.3.1]{MarsdenRatiu}.   The uniqueness of this one-parameter group can be shown by expressing the Poisson bracket in terms of the Poisson tensor $\Pi$, which in local coordinates gives
\[    \{a,b\}(x) = \sum_{i,j} \Pi^{ij}(x) \, (\partial_i a)(x) (\partial_j b)(x), \]
 and then noting that
\[   \frac{d}{dt} F^a_t(b)(x) = \sum_{i,j}  \Pi^{ij}(x) (\partial_i a)(x) \, (\partial_j  F^a_t(b))(x) \]   
has a unique local-in-time solution by standard methods in differential equations.
\end{proof}

Thus, the Poisson algebra arising from a compact Poisson manifold is a structure with a wonderful self-referential property: each element generates a one-parameter group of transformations preserving all the operations in this structure. 

We are now ready to prove a version of Noether's theorem.   Suppose that $A$ is a Poisson algebra and $a,b \in A$ generate one-parameter families.  
We say that $a$ generates \define{symmetries} of $b$, or $b$ is \define{conserved} by the family
generated by $a$, if
\[             F_t^a (b) = b       \qquad \forall t \in \R .  \]
Noether's theorem, formulated in terms of Poisson algebras, says the following:

\begin{thm} 
\label{thm:noether_1}
 Let $A$ be a Poisson algebra and suppose $a,b \in A$ generate one-parameter
 families.  Then $a$ generates symmetries of $b$ if and only if $b$ generates symmetries of
$a$.
\end{thm}

\begin{proof} 
The proof proceeds by a chain of equivalences.  Note that $a$ generates symmetries of $b$:
\[             F_t^a (b) = b       \qquad \forall t \in \R   \]
if and only if
\[               \frac{d}{dt} F_t^a (b) = 0    \qquad \forall t \in \R  \]
which in turn holds if and only if
\[      \left\{a, F_t^a (b) \right\} = 0    \qquad \forall t \in \R \]
which in turn holds if and only if 
\[            \{a,b \} = 0 . \]
By the antisymmetry of the bracket, i.e.\ $\{a,b\} = -\{b,a\}$, the last statement is true if and only if
\[            \{b,a\} = 0, \]
so running the chain of implications backwards with $a$ and $b$ switched we get
\[            F_t^b(a) = a \qquad \forall t \in \R ,   \]
meaning that $b$ generates symmetries of $a$.

The only step above that is not immediate is that 
\[     \{a,b \} = 0  \quad \implies \quad  \left\{a, F_t^a (b) \right\} = 0  \qquad \forall t \in \R. \]
To show this, assume $\{a,b\} = 0$.    Recall that $F_t^a(b)$ is the \emph{unique}
solution of Hamilton's equations
\[     \frac{d}{dt} F_t^a (b) =  \left\{a, F_t^a (b) \right\}   \qquad \forall t \in \R   \]
obeying the initial conditions
\[      F_0^a (b) = b .\]
Note that if we set $G(t) = b$ for all $t$, then $G(t)$ is another solution of Poisson 
equations:
\[     \frac{d}{dt} G(t) = 0 = \{a,b\} = \left\{a, G(t) \right\} \qquad \forall t \in \R   \]
obeying the same initial conditions.  By uniqueness we must have
\[     F_t^a(b) = b  \]
for all $t$, and thus
\[    \left\{a, F_t^a (b) \right\} = \{a,b\} = 0  \qquad \forall t \in \R ,  \]
as desired.
\end{proof}

The proof above has an arch-like structure, where the keystone of the arch 
is the antisymmetry of the bracket:
\begin{center}
$\{a,b\} = 0 \quad$ ($a$ generates symmetries of $b$) \break
$ \Updownarrow $ \break
$\{b,a\} = 0 \quad$  ($b$ generates symmetries of $a$)
\end{center}
So, in some sense the antisymmetry of the bracket contains the essence of Noether's
theorem.   What, however, is the meaning of the bracket's antisymmetry?   So far we
have simply \emph{posited} it in the definition of Poisson algebra, thus essentially
building Noether's theorem in from the start.

One possible answer is as follows.  Our original definition of Poisson algebra assumes the
bracket is linear in the second argument and antisymmetric.   If we instead assume it is bilinear---that is, linear in each argument---antisymmetry follows from assuming $\{a,a\} = 0$ for all $a$.  
The proof is easy:
\[   \{a,b\} + \{b,a\} =  \{a,a\} + \{a,b\} + \{b,a\} + \{b,b\} = \{a+b,a+b\} = 0.\]

If $a$ generates a one-parameter group, the equation $\{a,a\} = 0$ has a 
simple meaning: $a$ is conserved by the one-parameter group it itself generates.
We call this the \define{self-conservation principle}: \emph{each observable generates symmetries of itself}.   Most famously, energy is conserved by time translations.   Momentum in any direction is 
conserved by spatial translations in that direction.  Angular momentum about any axis is 
conserved by rotations around that same axis (but not necessarily other axes), and so on.  

In Theorem \ref{thm:noether_2} we derive a very general form of Noether's theorem from the self-conservation principle.   First we broaden the scope of investigation by bringing in quantum mechanics.

\section{Quantum mechanics}
\label{sec:quantum}

As we have seen, elements of a Poisson algebra play a dual role: they are both observables and generators of one-parameter families of transformations.   To accomodate this, a Poisson algebra is a hybrid structure, with the commutative algebra of observables and the Lie algebra of generators
tied together by the Leibniz law:
\[           \{ a, bc \} = \{a,b\} c + b \{a,c\} .  \]

The mathematics becomes more tightly unified in quantum mechanics.   Now we 
where we drop the commutativity of multiplication and decribe observables using  
a \define{real associative algebra}: a real vector space $A$ with a multiplication obeying
\[        a(bc) = (ab) c  \]
\[        a(\beta b + \gamma c) = \beta ab + \gamma ac  \qquad
       (\alpha a + \beta b)c  = \alpha a c + \beta b c \]
for all $a,b,c \in A$, $\alpha, \beta, \gamma \in \R$.   If we define the \define{commutator} $[-,-] \maps A \times A \to A$ by 
\[   [a,b] = ab - ba, \]
then we get the Lie algebra laws
\[          [a, [b,c]] =  [[a, b], c ] + [b, [a,c]]  \qquad [a,b] = -[b,a] \]
\[       [ a, \beta b + \gamma c ] = \beta [a,b] + \gamma [a,c]   \]
and the Leibniz law
\[          [a, bc] = [a,b] c + b [a,c] \]
\emph{for free}.  They follow from the associative algebra laws!

But there's a catch: not all elements of $A$ are observables, and not all are generators.  
To accomodate these two roles in one structure we typically require that $A$ be a \define{real $\ast$-algebra}, meaning an associative algebra with a real-linear operation $\ast \maps A \to A$ that is an \define{anti-homomorphism}:
\[         (ab)^\ast = b^\ast a^\ast  \qquad  \forall a, b \in A \]
 and \define{involutory}:
\[         (a^\ast)^\ast = a  \qquad \forall a \in A. \]
We then decree that
\begin{itemize}
\item generators are \define{skew-adjoint} elements of $A$, obeying $a^* = -a$.
\item observables are \define{self-adjoint} elements of $A$, obeying $a^* = a$.
\end{itemize}
To justify this, note that in a real $\ast$-algebra $A$, the set of skew-adjoint elements
\[    L = \{a \in A \colon \; a^* = -a \}  \]
is closed under the commutator $[a,b] = ab - ba$, and it is a Lie algebra, as one would expect 
of generators.   The set of self-adjoint elements
\[    O = \{a \in A \colon \; a^* = a \} \]
is closed under the \define{Jordan product}
\[          a \circ b =  \frac{1}{2}(ab + ba) \]
and it forms a \define{Jordan algebra}, meaning that 
\[      (a \circ a) \circ (a \circ b) = a \circ ((a \circ a) \circ b) \qquad
         a \circ b = b \circ a \]
\[        a \circ (\beta b + \gamma c ) = \beta (a \circ b) + \gamma (a \circ c) \]
for all $a,b,c \in A$, $\alpha, \beta, \gamma \in \R$.   We say more about why observables should
form a Jordan algebra in Section \ref{sec:jordan}.  The factor of $\frac{1}{2}$ is traditional here: it ensures that if $A$ has a multiplicative unit $1$ then $1 \circ a = a$.

Whenever $A$ is a real $\ast$-algebra we have
\[              A = O \oplus L  \]
because we can write any element $a \in A$ uniquely as the sum of self-adjoint
and skew-adjoint parts:
\[           a = \textstyle{ \frac{1}{2}(a+a^*) + \frac{1}{2}(a-a^*)  } .\]
This is an interesting situation: while observables and generators are packaged into a single algebraic structure, they are distinct.  

To probe this in a bit more detail, let us compare complex, real, and quaternionic quantum mechanics.  Quantum mechanics can be done using Hilbert spaces over all three of the associative normed division algebras $\R, \C$ and $\H$, and all three forms of quantum mechanics play a role in known physics \cite{Baez}.  However, they behave quite differently when it comes to the relation between observables and generators.

In complex quantum mechanics, the distinction between observables and generators is a mere formality.   Here we commonly take $A$ to be the algebra of bounded linear operators on a complex Hilbert space.  In this case, we can turn any self-adjoint bounded linear operator into a skew-adjoint one simply by multiplying it by $i$.  This sets up an isomorphism of real vector spaces
\[         \begin{array}{rcl} 
\psi \maps O &\to & L \\
                 a &\mapsto & ia 
\end{array}
\]

The distinction between observables and generators becomes significant in real quantum mechanics, where $A$ is the algebra of bounded linear operators on a \emph{real} Hilbert space.  
There is no way to multiply an observable by $i$ and get a generator.  Indeed, suppose $A$ is the algebra of linear transformations of $\R^n$, treated as real Hilbert space with its usual inner product.   Then $O$ has dimension $n(n+1)/2$, while $L$ has dimension $n(n-1)/2$.  These spaces are not isomorphic when $n > 0$: there are ``more observables than generators''.  In fact, there is no nonzero linear map from $O$ to $L$, or vice versa, that is invariant under the relevant group of symmetries, namely the orthogonal group $\mathrm{O}(n)$.  

Alternatively, consider the case where $A$ is the algebra of bounded linear operators on a quaternionic Hilbert space.   Now we can multiply an observable by $i$, but the result is not a generator.  Indeed, if the quaternionic Hilbert space is $n$-dimensional, $O$ has dimension $2n-n^2$ while $L$ has dimension $2n^2+n$.  These spaces are again not isomorphic when $n > 0$, but now there are ``more generators than observables''.   And again, there is no nonzero linear map between $O$ and $L$ that is invariant under the relevant symmetry group, namely the quaternionic unitary group $\mathrm{Sp}(n)$. 

All this illustrates a clear superiority of complex quantum mechanics over the real
and quaternionic versions.  Only in the complex case can we identify observables with
generators of one-parameter groups!  Thus, only in the complex case can we state
Noether's theorem in this form:
\begin{center}
The one-parameter group generated by $a$ preserves the observable $b$.
\[ \Updownarrow \]
The one-parameter group generated by $b$ preserves the observable $a$.
\end{center}
It is the number $i \in \C$ that lets us turn observables into generators, and vice versa.

In what follows we make this idea precise in a more general context.   First we prove
an easy result, showing that the complex numbers are \emph{sufficient} to unify
generators and observables in a framework where Noether's theorem holds.   In 
Section \ref{sec:jordan} we discuss results showing that the complex
numbers are also \emph{necessary}, given some technical side-conditions.

In complex quantum mechanics we can treat generators and observables in a unified
fashion using a `complex $\ast$-algebra'.   A \define{complex associative algebra} is a complex vector space $A$ with a multiplication such that the associative algebra laws hold 
\[        a(bc) = (ab) c  \]
\[        a(\beta b + \gamma c) = \beta ab + \gamma ac  \qquad
       (\alpha a + \beta b)c  = \alpha a c + \beta b c \]
for all $a,b,c \in A$ and $\alpha, \beta, \gamma \in \C$.   A \define{complex $\ast$-algebra} 
is a complex associative algebra with an operation $\ast \maps A \to A$ that is an involutory antihomomorphism and also \define{antilinear}:
\[       (\alpha a)^\ast = \overline{\alpha} a^\ast  \]
for all $\alpha \in \C, a \in A$.     

Any complex $\ast$-algebra $A$ has an underlying real $\ast$-algebra, so we can define a real vector space of skew-adjoint elements:
\[    L = \{a \in A \colon \; a^* = -a \}  \]
which forms a Lie algebra under the commutator, and a real vector space of self-adjoint elements:
\[    O = \{a \in A \colon \; a^* = a \} \]
which forms a Jordan algebra under the Jordan product.  But now we also have an isomorphism
of real vector spaces:
\[         \begin{array}{rcl} 
\psi \maps O &\to & L \\
                 a &\mapsto & ia .
\end{array}
\]
Given a complex $\ast$-algebra $A$, we say $a \in O$ \define{generates a one-parameter family} if for each $b \in O$ there exists a unique solution $F_t^a(b)$ to \define{Heisenberg's equation}
\[      \frac{d}{dt} F_t^a(b) = [ia, F_t^a(b)]   \qquad \forall b \in B \]
obeing the initial conditions
\[    F_0^a(b) = b. \]
As usual, we say $a$ generates \define{symmetries} of $b$ if
\[     F_t^a(b) = b \qquad \forall t \in \R .\]
We then have this version of Noether's theorem:

\begin{thm}
\label{thm:noether_2}
Let $A$ be a complex $\ast$-algebra and suppose $a,b \in O$ generate one-parameter
families.  Then $a$ generates symmetries of $b$ if and only if $b$ generates symmetries
of $a$.
\end{thm}

\begin{proof}
The proof is formally just like that of Theorem \ref{thm:noether_1} for Poisson algebras.  
Note that $a$ generates symmetries of $b$ if and only if
\[               \frac{d}{dt} F_t^a (b) = 0    \qquad \forall t \in \R , \]
which holds if and only if
\[       [a, F_t^a (b)] = 0    \qquad \forall t \in \R \]
which in turn holds if and only if  $[a,b] = 0$.  This is true if and only if $[b,a] = 0$,
so running the argument in reverse we see that $b$ generates symmetries of $a$.

The only nontrivial step is that 
\[    [a,b] = 0  \quad \implies \quad  [a, F_t^a (b)] = 0  \qquad \forall t \in \R. \]
However, if we assume $[a,b] = 0$, the fact that $F_t^a(b)$ is the unique
solution of the Heisenberg equations
\[     \frac{d}{dt} F_t^a (b) =  [ia, F_t^a (b)]   \qquad \forall t \in \R   \]
obeying the initial conditions $F_0^a (b) = b$, 
together with the fact that the constant function $G(t) = b$ is another solution with
the same initial conditions, implies that $F_t^a(b) = b$.  We thus obtain
\[    [a, F_t^a (b) ] = [a,b] = 0  \qquad \forall t \in \R \]
as desired.
\end{proof}

The theorem above leaves open the question of which observables generate one-parameter
families.   This is a question about existence and uniqueness of solutions of differential equations, so we need a bit of analysis to solve it.   The nicest complex $\ast$-algebras are the C*-algebras.  In a C*-algebra, \emph{every} observable generates a one-pameter family. 

A \define{C*-algebra} is a complex $\ast$-algebra $A$ equipped with a norm making it into a Banach space (i.e.\ a complete normed vector space) and obeying 
\[             \|ab\| \le \|a\| \, \|b\|  \qquad       \|aa^*\| = \|a\|^2 .\]
for all $a, b \in A$.   The following result is well-known, so we merely sketch the proof: 

\begin{thm}
\label{thm:C*-algebra}
If $A$ is a C*-algebra then every element of $O =  \{a \in A \colon \; a^* = a \}$ generates a one-parameter group of transformations that preserve all the C*-algebra structure.
\end{thm}

\begin{proof}
If $a \in O$ then for each $t \in \R$ we can define
\[   e^{ita} = \sum_{n=0}^\infty  \frac{(ita)^n}{n!} \in A  \]
where the power series converges in the norm topology because
\[  \left\|  \frac{(ita)^n}{n!}  \right\| \le  \frac{t^n \|a\|^n}{n!}. \]
For any $b \in O$ we can define
\[     F_t^a(b) = e^{ita} b e^{-ita}. \]
One can show by manipulating convergent formal power series that $F_t^a(b) \in O$ for
all $t \in \R$, that $F_t^a(b)$ obeys Heisenberg's equation with the initial conditions
$F_t^a(b) = b$, and that $F_t^a \maps O \to O$ is a one-parameter group:
\[        \begin{array}{ccl}           F^t_0 &=& \mathrm{id}   \\ 
              F^f_{s+t} &=&  F^f_s \circ F^f_t  \qquad \forall s, t \in \R   .\end{array} \]
For each $b$ one can show that $F_t^a(b)$ is the unique solution of Heisenberg's equations with the given initial conditions.   Moreover one can show by calculations that the transformations $F_t^a$ preserve all the C*-algebra structure:
\[       \begin{array}{ccl}
F_t^a(\beta b + \gamma c) &=& \beta F_t^a(b) + \gamma F_t^a(c)  \\ 
F_t^a(bc) &=& F_t^a(b) F_t^a(c)  \\  
F_t^a(b^*) &=& F_t^a(b)^*   \\  
\|F_t^a(b)\| &=& \|b\| 
\end{array}
\]
for all $b,c \in A$, $\beta,\gamma \in \R$.
\end{proof}

A huge amount of work on quantum physics has been done using C*-algebras, because they are a very flexible framework that still has close ties to the more traditional complex Hilbert space approach \cite{Arveson,BratteliRobinson,Emch,Haag}.   The so-called `C*-axiom' $\|a\|^2 = \|a^*a\|$ is very powerful.   Among other things, it provides a concrete answer to the question of the physical \emph{meaning} of the norm.   This works as follows.  Any element $a$ in a unital algebra $A$ has a \define{spectrum} 
\[           \sigma(a) = \{z \in \C : \; a - z 1 \textrm{ is not invertible} \} . \]
If $A$ is a C*-algebra and $a \in A$ is self-adjoint then $\sigma(a) \subseteq \R$, and 
\[        \|a\| = \sup \{ |z| : \;   z \in \sigma(a) \}  .\]
We think of $\sigma(a)$ as the set of values the observable $a$ can assume.  Thus, $\|a\|$ is simply the supremum of $|z|$ where $z$ ranges over all values that $a$ can take on.  Furthermore, one can define `states' on a C*-algebra, and given an observable $a$ and a state $\omega$ we obtain a probability measure on the real line that is supported on the spectrum of $a$.  This gives the probability that $a$ takes on values in any measurable set $S \subseteq \R$ in the state $\omega$.   We describe this in more detail in a more general framework in Section \ref{sec:jordan}.  

To dig deeper into the meaning of Noether's theorem, in the next sections we separately study the two halves of a $\ast$-algebra: Lie algebras, which describe generators, and Jordan algebras, which describe observables.  In the end we shall see that a special kind of Jordan algebra automatically gives rise to a Lie algebra---and under certain conditions connected to Noether's theorem, we can combine the Jordan algebra with this Lie algebra to obtain a C*-algebra.

To lay the groundwork, let us see what happens when we express all the structure of a real or complex $\ast$-algebra in terms of observables and generators.    When we do this for a real
$\ast$-algebra we get a rather elaborate structure.   When we do it for a complex $\ast$-algebra
we can simplify the answer, since we can use multiplication by $i$ to express generators in terms
of observables, or the other way around.

\begin{thm} 
\label{thm:real_star-algebra}
Given a real $\ast$-algebra $A$, let 
\[  L = \{a \in A \colon \; a^* = -a \}  ,  \qquad O = \{a \in A \colon \; a^* = a \}  \]
and define
\[    \{a,b\} = \frac{1}{2}(ab - ba) \qquad a \circ b = \frac{1}{2}(ab+ba)  \]
Then the following conditions hold:
\begin{enumerate}
\item The operations
\[    \{-,-\} \maps L \times L \to L    \qquad \{-,-\} \maps L \times O \to O \]
\[    \{-,-\} \maps O \times L \to O    \qquad \{-,-\} \maps O \times O \to L \]
are bilinear and obey $\{a,b\} = -\{b,a\}$
for all $a,b$ in either $L$ or $O$.
\item  The operations
\[    \circ \maps L \times L \to O    \qquad \circ \maps L \times O \to L \]
\[    \circ \maps O \times L \to L    \qquad \circ \maps O \times O \to O \]
are bilinear and obey $a \circ b = b \circ a$
for all $a,b$ in either $L$ or $O$.
\item For all $a,b,c$ in either $L$ or $O$,
\[    \{a, b \circ c\} = \{a, b\} \circ c + b \circ \{a,c\} .   \]
\item  For all $a,b,c$ in either $L$ or $O$,
\[    (a \circ b) \circ c - a \circ (b \circ c) = \{a,\{b,c\}\} - \{\{a,b\},c\} .  \]
\end{enumerate}
Conversely, given any pair of vector spaces $L,O$ and any operations obeying conditions 1--4, $A = L \oplus O$ becomes a real $\ast$-algebra if we define
\[       ab = a \circ b + \{a,b\}    \]
and define a linear map $\ast \colon A \to A$ by $a^* = a$ for $a \in O$, $a^* = -a$ for $a \in L$.
\end{thm}

\begin{proof}
The proof is a calculation.  In the forward direction, we simply check that given a real $\ast$-algebra, conditions 1--4 hold.   In the reverse direction, to check the associative law
\[            (ab)c = a(bc)   \]
for all $a,b,c \in O \oplus L$, we expand it using $ab = a \circ b + \{a,b\} $ to obtain
\[         (a \circ b) \circ c + \{a \circ b,c\} + \{a,b\} \circ c + \{\{a,b\},c\} = \]
\[          a \circ (b \circ c) + \{a, b \circ c\} + a \circ \{b,c\} + \{a,\{b,c\}\}.   \]
To check this equation we use item 3, the anticommutative law for $\{-,-\}$, and the commutative law for $\circ$ to simplify it to
\[      (a \circ b) \circ c - a \circ (b \circ c) = \{a,\{b,c\}\} - \{\{a,b\},c\}  ,\]
which is item 4.   The law $(ab)^* = b^* a^*$ follows from the definitions,  the anticommutative law for $\{-,-\}$, and the commutative law for $\circ$.  The other $\ast$-algebra laws are easier to check.
\end{proof}

Remarkably, in this theorem we do not assume from the start that $O$ is a Jordan algebra or $L$ is a Lie algebra to prove that $O \oplus L$ is a real $\ast$-algebra.   The most complicated of the Jordan and Lie algebra laws, namely the \define{Jordan identity}
\[   (a \circ a) \circ (a \circ b) = a \circ ((a \circ a) \circ b)  \]
and the Jacobi identity
\[   \{a,  \{b,c\}\} = \{\{a, b\}, c\} + \{b, \{a,c\}\}, \]
actually \emph{follow} from conditions 1--4.   Moreover, conditions 3 and 4 cannot be derived from the Jordan and Lie algebra laws alone.    This raises a question: what do these conditions mean?   We address this in Section \ref{sec:jordan}.

Condition 1 in Theorem \ref{thm:complex_star-algebra} implies this purely algebraic version of Noether's theorem:
\[   \{a,b\} = 0 \iff \{b,a\} = 0  \qquad \forall a,b \in O.\]
Thus, we can say that a bracket operation $\{-,-\} \maps O \times O \to L$ obeying this version of Noether's theorem is a prerequisite for $O$ to become the space of self-adjoint elements of a real $\ast$-algebra.   

The result for complex $\ast$-algebras is simpler:

\begin{thm} 
\label{thm:complex_star-algebra}
Given a complex $\ast$-algebra $A$, let $O = \{a \in A : a^* = a \}$ and define
\[    \{a,b\} = \textstyle{\frac{i}{2}}(ab - ba) \qquad a \circ b = \frac{1}{2}(ab+ba)  \]
for all $a,b \in O$.
Then
\begin{enumerate}
\item The operation $\{-,-\} \maps O \times O \to O$ is bilinear and obeys
\[     \{a,b\} = -\{b,a\}  \qquad \forall a,b \in O .\] 
\item  The operation $\circ \maps O \times O \to O$ is bilinear and obeys
\[     a \circ b = b \circ a   \qquad \forall a,b \in O . \]
\item For all $a,b,c \in O$,
\[    \{a, b \circ c\} = \{a, b\} \circ c + b \circ \{a,c\} .   \]
\item  For all $a,b,c$ in $O$,
\[    (a \circ b) \circ c - a \circ (b \circ c) = \{\{a,b\},c\} - \{a,\{b,c\}\}.  \]
\end{enumerate}
Conversely, given a real vector space $O$ and operations obeying conditions 1--4, the complex
vector space $A = \C \otimes O$ becomes a complex $\ast$-algebra in a unique way such that
\[       ab = a \circ b - i\{a,b\}   \]
and
\[    (a + ib)^* = a - ib   \]
for all $a,b \in O$.   
\end{thm}

\begin{proof}
This follows from Theorem \ref{thm:real_star-algebra} by noting that for a complex $\ast$-algebra we can use the linear bijection
\[         \begin{array}{rcl} 
\psi \maps O &\to & L \\
                 a &\mapsto & ia 
\end{array}
\]
to express all the structure in terms of $O$, and rewrite $A = L \oplus O$ as 
$\mathbb{C} \otimes O = \{ a + ib : \; a,b \in O \}$.   Beware: the 
bracket $\{-,-\}$ is defined with an extra factor of $i$ in the present theorem, so 
condition 4 here differs in sign compared to condition 4 in Theorem \ref{thm:real_star-algebra}.
\end{proof}

The above result is implicit in Alfsen and Shultz's theorem on `dynamical correspondences' \cite{AlfsenShultz98}, and in Section \ref{sec:jordan} we use it to sketch a proof of their theorem.  The reader should also compare work on Jordan--Lie--Banach algebras \cite{Emch2,GrginPetersen,HalvorsonClifton,Landsmaan}. 

\section{Lie algebras}
\label{sec:lie}

In Section \ref{sec:classical} we described a link between Noether's theorem and the
`self-conservation principle', namely that each observable generates symmetries of itself.  Using
this idea we can present a version of Noether's theorem with minimal hypotheses.   Looking at the proof of Theorems \ref{thm:noether_1} and \ref{thm:noether_2}, we see they use nothing about the multiplication of observables or the Jacobi identity for the bracket.  Thus we can eliminate these, and suppose we merely have vector space with a bracket operation obeying a few conditions:

\begin{thm}
\label{thm:noether_3}
Suppose $L$ is a vector space with a bilinear map $\{-,-\} \maps L \times L \to L$ such that:
\begin{itemize}
\item
For each $a,b \in L$ there is a unique one-parameter family of elements $F_t^a(b) \in L$
obeying 
\[     \frac{d}{dt} F_t^a(b) = \{a, F_t^a(b)\}  \quad \forall t \in \R \]
with initial conditions
\[      F_0^a(b) = b  .\]
\item Each $a \in L$ generates symmetries of itself:
\[       F_t^a(a) = a  \qquad \forall t \in \R.  \]
\end{itemize}
Then for any pair $a, b \in L$, this version of Noether's theorem holds:
\[ 
\begin{array}{rcl}    F^a_t (b) &=& b \quad \forall t \in \R  \\
& \Updownarrow &  \\
  F^b_t (a) &=& a \quad \forall t \in \R .
\end{array}
\]
That is, $a$ generates symmetries of $b$ if and only if $b$ generates symmetries of $a$.
\end{thm}

\begin{proof}
 First note that the hypotheses imply
\[    \{a,a\} = \frac{d}{dt} F_t^a(a) \Big|_{t = 0} = \frac{da}{dt} = 0 \]
for all $a \in A$.   Since the bracket is bilinear this implies that the bracket is antisymmetric.
From this point on, the proof is exactly like that of Theorem \ref{thm:noether_1}.   
\end{proof}

This theorem amounts to dropping \emph{observables} entirely and stating Noether's theorem
purely in terms of \emph{generators}.  In practice, the vector spaces obeying the conditions of Theorem \ref{thm:noether_3} are almost invariably Lie algebras.  Thus, to dig deeper, we need to ask why Lie algebras are important in physics.  

In one sense the answer to this question is simple enough: Lie algebras tend to come from Lie groups, and symmetries in physics tend to form Lie groups.  Hilbert's fifth problem goes further and asks which topological groups are Lie groups.  After decades of work, mathematicians arrived at a nice answer.

\begin{thm}
\label{thm:hilbert_5th}
Suppose $G$ is a connected locally compact topological group for which the identity has a neighborhood containing no subgroup except the trivial group.   Then $G$ is a Lie group.
If $G$ is a Lie group then its tangent space at the identity has the structure of a finite-dimensional Lie algebra.  Conversely, any finite-dimensional Lie algebra arises in this way from some topological group $G$ of this form, which is unique up to isomorphism if we also require that it is simply connected.
\end{thm}

\begin{proof}
The first statement follows from results of Gleason and Yamabe building on a long line
of previous work \cite{Tao}.  The rest is a combination of Lie's first, second and third theorems.
\end{proof}

Note that the assumptions on the group $G$ above do not even mention the real numbers, and yet we conclude that $G$ is a manifold!  If we assume the symmetries of a physical system form a topological group with these properties, we get a finite-dimensional Lie algebra and---we shall see---a version of Noether's theorem for free.

But infinite-dimensional Lie algebras are also important in physics, so to state this version of Noether's theorem, let us start with a larger class of Lie algebras.   A \define{Banach--Lie algebra} is a Banach space $L$ that is also a Lie algebra where the bracket obeys
\[          \|[a,b]\| \le C \|a\| \|b\|      \qquad a, b \in L   \]
for some $C > 0$.  By rescaling the norm we can assume $C = 1$.  Any finite-dimensional Lie algebra can be given a norm making it a Banach--Lie algebra.  Also, the algebra of bounded linear operators on any Banach space $V$ becomes a Banach--Lie algebra using the operator norm
\[    \|a\| = \sup_{\psi \in V, \psi \ne 0} \frac{\|a \psi\|}{\|\psi\|}  . \]

\begin{thm}
\label{thm:noether_4} Suppose $L$ is a Banach--Lie algebra.   Then for each $a,b \in L$ there is a unique one-parameter family of elements $F_t^a(b) \in L$ obeying 
\[     \frac{d}{dt} F_t^a(b) = [a, F_t^a(b)]  \quad \forall t \in \R \]
with initial conditions
\[      F_0^a(b) = b  .\]
Moreover $F_t^a \maps L \to L$ is one-parameter group of bounded linear transformations preserving the Lie bracket.  For any pair $a,b \in L$, this version of Noether's theorem holds:
\[ 
\begin{array}{rcl}    F^a_t (b) &=& b \quad \forall t \in \R  \\
& \Updownarrow &  \\
  F^b_t (a) &=& a \quad \forall t \in \R .
\end{array}
\]
\end{thm}

\begin{proof}
If $L$ is a Banach--Lie algebra and $a \in L$, the linear transformation $T \maps L \to L$ given by
\[     T(b) = [a,b]     \]
is bounded.  By Picard's existence and uniqueness theorem for the solution of ordinary differential equations, whose usual proof also works for Banach-space valued functions, for each $b \in L$ there is a unique solution $F_t^a(b)$ of the differential equation 
\[     \frac{d}{dt} F_t^a(b) = [a, F_t^a(b)] = T(F_t^a(b)) \]
with initial conditions 
\[     F_0^a(b) = b .  \]   
An explicit solution is given by $F_t^a(b) = \exp(tT)(b)$ where the exponential is defined by a convergent power series.  Thus, $F_t^a \maps L \to L$ is one-parameter group of bounded linear transformations.   We can show these transformations preserve the Lie bracket by noting that for all  $t \in \R$ and $b,c \in L$ we have
\[  \begin{array}{ccl}
\displaystyle{  \frac{d}{dt} [F_t^a(b), F_t^a(c)] }
  &=& [[a, F_t^a(b)], F_t^a(b)] + [F_t^a(b), [a,F_t^a(c)]] \\
  &=& [a, [F_t^a(b), F_t^a(c)]]  
  \end{array}
\]
while 
\[   \frac{d}{dt}  F_t^a([b,c]) = [a,  F_t^a([b,c])] .\]
Thus, if
\[    u(t) = F_t^a([b,c]) -  [F_t^a(b), F_t^a(c)] \]
we have
\[  \frac{d}{dt}u(t) = [a,u(t)] \qquad \forall t \in \R. \]
Since $u(0) = 0$, the uniqueness result just mentioned implies that
$u(t) = 0$ for all $t \in \R$.  Thus, the transformations $F_t^a$ preserve the Lie bracket:
\[    F_t^a([b,c]) =  [F_t^a(b), F_t^a(c)]   \]
for all $t \in \R$, $a,b,c \in L$.

To show that Noether's theorem holds, by Theorem \ref{thm:noether_3} it suffices to
show 
\[     F_t^a(a) = a  \qquad  \forall t \in \R, a \in L .\]
This follows from the uniqueness result, since 
\[    \frac{d}{dt} a = 0 = [a, a]  .   \qedhere \]
\end{proof}

Banach--Lie algebras arise naturally from a class of groups called Banach--Lie groups \cite{deLaHarpe}.  However, not every Banach--Lie algebra gives rise to such a group \cite{vanEstKorthagen}.  Every Banach--Lie algebra gives a \define{local Banach--Lie group}: a Banach manifold $G$ containing an element $1$, with partially defined  smooth group operations with open domains of definition, that obey the group laws when defined.   Conversely, any local Banach--Lie group has a Banach--Lie algebra.

Since a version of Noether's theorem holds automatically for any Banach--Lie algebra, and any such thing gives a local Banach--Lie group, we would be remiss not to use this fact to shed more light on Noether's theorem.   

Suppose $L$ is a Banach--Lie algebra and let $G$ be the corresponding local Banach-Lie group.  (The reader will lose nothing by considering the case where $G$ is a Lie group.)    Suppose $a,b \in L$.   From the proofs of Theorems \ref{thm:noether_3} and \ref{thm:noether_4} we see that in this context, Noether's theorem
\begin{center}
$a$ generates transformations that leave $b$ fixed $\iff$ \break $b$ generates transformations that leave $a$ fixed
\end{center}
boils down to the statement
\[     [a,b] = 0 \iff [b,a] = 0 .\]
Given $a,b \in L$ we have smooth curves $\exp(sa), \exp(tb)$ in $G$, defined for small $s$ and $t$, and
we have
\[          [a,b] = \left. \frac{d^2}{ds dt} \exp(sa) \exp(tb) \exp(-sa) \exp(-ta) \right|_{s,t=0} \]
It follows that $[a,b] = 0$ if $\exp(sa)$ and $\exp(tb)$ commute for sufficiently small
$s,t$.  By the Baker--Campbell--Hausdorff formula the converse is true too.  

Therefore, Noether's theorem, phrased solely in terms of generators, follows from the fact that
\begin{center}
$g$ commutes with $h$ $\iff$ $h$ commutes with $g$.  
\end{center}
It is simply an `infinitesimal' version of this obvious fact!  Similarly, the self-conservation principle follows from the fact that
\begin{center}
$g$ commutes with $g$.
\end{center}

\section{Jordan algebras}
\label{sec:jordan}

Having reduced Noether's theorem to a triviality by working only with generators, we see that its nontrivial content must involve observables.   Thus, we turn to Jordan algebras.  These were introduced by Pascual Jordan \cite{Jordan}, who studied them further with  von Neumann and Wigner in an attempt to isolate the algebraic properties of observables in quantum mechanics \cite{JordanvonNeumannWigner}.    By now much is known about them \cite{Chu2011,Hanche-OlsonStormer,Jacobson,McCrimmon,Upmeier}.  Still, they remain less widely understood than Lie algebras, so before examining their connection to Noether's theorem we start by explaining them.  We shall see that a certain class of Jordan algebras, the `unital JB-algebras', are a powerful framework for studying the connection between Jordan algebras and the role of the complex numbers in quantum mechanics.

Recall that a Jordan algebra is a real vector space $O$ equipped with an operation $\circ \maps O \times O \to O$ obeying
\[        (a \circ a) \circ (a \circ b) = a \circ ((a \circ a) \circ b) \qquad
         a \circ b = b \circ a \]
\[        a \circ (\beta b + \gamma c ) = \beta (a \circ b) + \gamma (a \circ c) \]
for all $a,b,c \in O$, $\alpha, \beta, \gamma \in \R$.   These laws were found by listing some of the properties of the Jordan product $a \circ b = \frac{1}{2}(ab + ba)$ of $n \times n$ self-adjoint complex matrices.  

The law $(a \circ a) \circ (a \circ b) = a \circ ((a \circ a) \circ b)$ is mysterious: it demands an explanation.   If we use the commutative law to rewrite it this way:
\[               (a \circ a) \circ (b \circ a) = ((a \circ a) \circ b) \circ a \]
we see it is a special case of the associative law.    But there is perhaps a better way to think about it.   One can show, through a tiresome inductive argument, that any Jordan algebra is \define{power-associative}: given any element $a$, the $n$-fold product $a \circ \cdots \circ a$ is independent of how we parenthesize it.  Thus we can write this element as $a^n$ without ambiguity.   Furthermore we can prove that 
\[             a^m  \circ (a^n \circ b) = a^n \circ (a^m \circ b)  \qquad \forall m, n \ge 1    \]
for all $a,b$ in a Jordan algebra \cite[Theorem 5.2.2]{McCrimmon}.    The mysterious law in the definition of Jordan algebra is just the case $m = 2, n = 1$.  Thus, we can equivalently define a Jordan algebra to be a vector space with a commutative power-associative bilinear operation $\circ$ having this property: for any element $a$, all the operations of multiplication by powers $a^n$ commute.

It is difficult to directly explain the physical meaning of $a \circ b$, since there is not a general procedure for measuring $a \circ b$ given a way to measure $a$ and a way to measure $b$.
However, in a Jordan algebra we have
\[    a \circ b = \textstyle{\frac{1}{2}} \big((a + b) \circ (a + b) - a \circ a - b \circ b\big) \]
for all $a,b$, so we can understand the meaning of $a \circ b$ if we can understand linear combinations of observables and also the \define{Jordan square} of an observable, $a^2 = a \circ a$.

For this we need to define a concept of `state' for a Jordan algebra, and connect
observables to states.  This can be done, not for all Jordan algebras, but for a large class. 
We say a Jordan algebra \define{formally real} if 
\[    a_1^2 + \cdots + a_n^2 = 0 \implies a_1, \dots, a_n = 0  \]
for all $a_1, \dots, a_n \in O$.   This captures our intuition that the sum of squares of real-valued observables can vanish only if each observable vanishes separately.   

The condition of being formally real is very powerful.   Jordan, von Neumann and Wigner \cite{JordanvonNeumannWigner} showed that in the finite-dimensional case, this condition
implies the mysterious law $(a \circ a) \circ (a \circ b) = a \circ ((a \circ a) \circ b)$ so long as the operation $\circ$ is also commutative, power-associative and bilinear.  They also classified the finite-dimensional formally real Jordan algebras.  They began by proving that any such algebra is a direct sum of `simple' ones.  A Jordan algebra is {\bf simple} when its only ideals are $\{0\}$ and $A$ itself, where an {\bf ideal} is a vector subspace $B \subseteq A$ such that $b \in B$ implies $a \circ b \in B$ for all $a \in A$.   Then they proved every simple finite-dimensional formally real Jordan algebra is isomorphic to one on this list:
\begin{itemize}
\item The algebra $\h_n(\R)$ of $n \times n$ self-adjoint real 
matrices with the product $a \circ b = \frac{1}{2}(ab + ba)$.
\item The algebra $\h_n(\C)$ of $n \times n$ self-adjoint complex 
matrices with the product $a \circ b = \frac{1}{2}(ab + ba)$.
\item The algebra $\h_n(\H)$ of $n \times n$ self-adjoint quaternionic 
matrices with the product $a \circ b = \frac{1}{2}(ab + ba)$.
\item The algebra $\h_3(\O)$ of $3 \times 3$ self-adjoint octonionic 
matrices with the product $a \circ b = \frac{1}{2}(ab + ba)$.
\item The algebras $\R^n \oplus \R$ with product $(x,t) \circ (x', t') =
(t x' + t' x, x \cdot x' + tt')$.
\end{itemize}
All these are \define{unital}: they have an element $1$ that obeys $1 \circ a = a$ for all $a$.  Thus, all finite-dimensional formally real Jordan algebras are unital.

Unital formally real Jordan algebras are important because we can define `states' of any physical system whose observables form such an algebra.  The first step is to note  that in a formally real Jordan algebra $O$ we can define $a \ge b$ to mean that $a-b$  is a sum of Jordan squares.   This relation is a \define{partial order}, since it clearly obeys the laws
\[         a \ge a \qquad a \ge b \;\& \; b \ge c \implies a \ge c \qquad a \ge b \; \& \; b \ge a \implies a = b \]
for all $a,b,c \in O$.   One can also see that 
\[        a,b \ge 0 \implies \alpha a + (1-\alpha) b \ge 0   \qquad \forall \alpha \textrm{ s.t. } 0 \le \alpha \le 1 \]  
\[        a \ge 0 \implies \alpha a \ge 0  \qquad  \forall \alpha > 0\]
for all $a,b \in O$.   We summarize the last two facts by saying that the set
\[         O^+ = \{a \in O : \; a \ge 0 \}  \]
is a \define{convex cone}.   Conversely, one can construct a formally real Jordan algebra from any convex cone whose interior is `symmetric' in a certain sense.  In the finite-dimensional case this is nicely explained in the text by Faraut and Koranyi \cite{FarautKoranyi}; for an infinite-dimensional generalization see Chu \cite{Chu2017}.  

Given a unital formally real Jordan algebra $O$, a `state' is a particular kind of linear functional sending  each observable $a \in O$ to its expected values.    A linear functional $\omega \maps O \to \R$ is called \define{positive} if
\[           a \ge 0 \implies \omega(a) \ge 0 . \]
A \define{state} is a positive linear functional $\omega \maps O \to \R$ that is \define{normalized}:
\[           \omega(1) = 1. \]
The states form a convex set: 
\[      \omega, \nu \textrm{ are states} \implies \alpha \omega + (1-\alpha) \nu \textrm{ is a state }   \qquad \forall \alpha \textrm{ with } 0 \le \alpha \le 1. \] 

There has been much work on `generalized probabilistic theories', where one starts with a 
convex set of states, or perhaps a cone with an `order unit' playing the role of $1 \in O^+$.  There are a number of interesting attempts to justify a focus on generalized probabilistic theories arising from finite-dimensional formally real Jordan algebras \cite{BarnumGraydonWilce, BarnumHilgert, Wilce11, Wilce18}.   To generalize this work to infinite-dimensional Jordan algebras we need a bit of analysis, so we use `JB-algebras'.   

A \define{JB-algebra} is a Jordan algebra $O$ equipped with a norm making it into a Banach space and obeying
\[    \|a \circ b\| \le \|a\| \|b\|  \qquad   \|a^2\| = \|a\|^2 \qquad  \|a^2\| \le \|a^2  + b^2\| \]
for all $a,b \in O$.    The last inequality implies that any JB-algebra is formally real.   Conversely, any \emph{finite-dimensional} unital formally real Jordan algebra can be given a norm making it JB-algebra---but this fails in the infinite-dimensional case \cite[Sec.\ 3.1]{Hanche-OlsonStormer}.  So, the concept of JB-algebra really comes into its own in the infinite-dimensional case.   

Given any state $\omega \maps O \to \R$ on a unital JB-algebra, and any element $a \in O$, there is a probability measure on the real line that describes what happens when one measures the observable $a$ in the state $\omega$.    This is the unique Borel measure $\mu_{a,\omega}$ such that 
\[             \omega(a^n) = \int_\R x^n \, d\mu_{a,\omega}(x)    \qquad \forall n \ge 0 .\]
This measure is supported on the \define{spectrum} of $a$, which is defined just as for C*-algebras:
\[           \sigma(a) = \{z \in \C : \; a - z 1 \textrm{ is not invertible} \} . \]
The spectrum of $a \in O$ is always a compact subset of the real line, and we have
\[        \|a\| = \sup \{ |z| : \;   z \in \sigma(a) \}  .\]
Thus, we can think of $\sigma(a)$ as the set of values the observable $a$ can assume, and
$\|a\|$ as the supremum of $|z|$ over all values $z$ that $a$ can assume.  Furthermore, this formula implies that the norm is not really an extra structure: if a unital Jordan algebra has a norm making it a JB-algebra, this norm is unique.   For details, see \cite[Prop.\ 2.4]{AlfsenShultzStormer} and the discussion that follows that proposition.

Another nice feature of JB-algebras is that they support a `functional calculus'.  This lets us apply continuous functions to elements of a unital JB-algebra.   For any $a \in O$
and any continuous function $f \maps \sigma(a) \to \R$ we can define $f(a) \in O$ such that
\[             \omega(f(a)) = \int_\R f(x) \, d\mu_{a,\omega}(x)    \]
for every state $\omega \maps O \to \R$.    

Now let us turn to Noether's theorem.  The self-adjoint elements of any C*-algebra form a JB-algebra if we set $a \circ b = \frac{1}{2}(ab+ ba)$.    As we saw in Theorems  \ref{thm:noether_2} and \ref{thm:C*-algebra}, C*-algebras have Noether's theorem `built in': every observable generates a one-parameter group of transformations preserving all the C*-algebra structure, and $a$ generates symmetries of $b$ if and only if $b$ generates symmetries of $a$.     Alfsen and Shultz \cite{AlfsenShultz98} compared the  situation for JB algebras.  They showed that a JB-algebra consists of a self-adjoint elements of a C*-algebra if and only if it obeys some conditions including a version of Noether's theorem.  We now discuss this result.

For the rest of this section let $O$ be a unital JB-algebra.   We begin by getting our hands on a Lie algebra $L$ associated to $O$.    In terms of physics, if $O$ is the Jordan algebra of observables of some system, $L$ will be its Lie algebra of generators.   

We start by defining a group.   The \define{order automorphisms} of $O$ are the invertible linear transformations $f \maps O \to O$ that preserve the partial order:
\[           f(a) \ge 0 \iff a \ge 0    \qquad \forall a \in O. \]
Order automorphisms are automatically bounded, and they form a Banach--Lie group $\Aut_{\ge}(O)$.     A bounded linear map $\delta \maps O \to O$ is an \define{order derivation} if
\[      \exp(t \delta) \in \Aut_\ge(O) \qquad \forall t \in \R .\]
where the exponential is defined using the usual power series.  Order derivations naturally form a vector space which we call $A$, and this is a Lie algebra under commutators \cite[Prop.\ 10]{AlfsenShultz98}.  This is the Lie algebra of the Banach--Lie group $\Aut_{\ge}(O)$.

Every order derivation can be written uniquely as a sum of two kinds, called `skew-adjoint' and `self-adjoint' \cite[Lemma 11]{AlfsenShultz98}.   The skew-adjoint order derivations are closed under commutators so they form a Lie algebra $L$ in their own right, a Lie subalgebra of $A$.  The self-adjoint ones correspond to elements of $O$, and these are not closed under commutators.  We thus have 
\[   A = O \oplus L \]
as vector spaces, but not as Lie algebras.

A \define{skew-adjoint} order derivation is a bounded linear map $\delta \maps O \to O$ obeying any one of these equivalent conditions:
\begin{enumerate}
\item $\delta$ is an order derivation and $\delta 1 = 0$.
\item $\delta$ is a derivation of the Jordan product:
\[     \delta (a \circ b) = \delta a \circ b + a \circ \delta b  \qquad \forall a,b \in O .\] 
\item For all $t \in \R$, $\exp(t \delta)$ preserves the Jordan product:
\[        \exp(t\delta)(a \circ b) = \exp(t \delta)(a) \circ \exp(t \delta)(b)   \qquad \forall a,b \in O \]
\item If $\omega$ is a state on $O$ then $\omega \circ e^{t\delta}$ is a state on $O$ for all $t \in \R$.
\end{enumerate}
The equivalence of these conditions follows from \cite[Lemma 9]{AlfsenShultz98}.  Using condition 2 it is easy to see that skew-adjoint order derivations are closed under commutators.  Thus they form a Lie subalgebra $L$ of the Lie algebra $A$ of all order derivations.

On the other hand, any element $H \in O$ gives an order derivation $\delta_H$ defined by
\[        \delta_H b = H \circ b    \qquad \forall b \in O \]
and such order derivations are called \define{self-adjoint}.    If $O$ consists of all the self-adjoint elements of a C*-algebra, any self-adjoint order derivation $\delta_H$ generates a one-parameter group as follows:
\[    \exp(s \delta_H) (b) = \exp(sH/2) \, b \, \exp(sH/2) \qquad  \forall b \in O, s \in \R .\]
This formula does not make sense in a general unital JB-algebra, since it relies on the associative  multiplication in the C*-algebra.  However, while expressions of the form $aba$ do not make literal sense in a Jordan algebra, McCrimmon \cite{McCrimmon} has emphasized that there is a perfectly good substitute.  For any $a \in O$ there is a linear map $U_a \maps O \to O$ given by
\[           U_a(b) = 2(a \circ (a \circ b)) - (a \circ a) \circ b  \qquad \forall a \in A  .\]
When $O$ is \define{special}---that is, when the Jordan product can be written as $a \circ b = \frac{1}{2} (a b + b a)$ using the multiplication in some possibly larger associative algebra---we have
\[          U_a(b) = a b a  .\]
Otherwise we can use $U_a(b)$ as a substitute for this expression.  Indeed, in any unital JB-algebra the one-parameter group generated by a self-adjoint derivation $\delta_H$ is given as follows:
\[      \exp(s \delta_H)(b) = U_{\exp(s H/2)}(b)    \qquad \forall b \in O, s \in \R \]
where  $\exp(s H/2)$ is defined using the functional calculus.  Concretely, we have
\[    \exp(s H/2) = \sum_{n = 0}^\infty \frac{(s H/2)^n}{n!} .   \]
Note that the individual terms in this sum make sense because $O$ is unital and power-associative; furthermore, the sum converges in the norm topology.

Let us see what order derivations amount to in some examples.  The skew-adjoint derivations are familiar, but the self-adjoint ones much less so.   

\begin{ex} 
\label{ex:matrix_algebras}
Let $\K = \R, \C$ or $\H$, let $A = M_n(\K)$ be the algebra of $n \times n$ matrices with entries in $\K$, and let
\[  \begin{array}{c} L = \{a \in A \colon \; a^* = -a \}  , \\
 O = \{a \in A \colon \; a^* = a \} .
 \end{array}
\]
Then $O$ is a JB-algebra with the Jordan product $a \circ b = \frac{1}{2} (ab + ba)$ and the operator norm
\[    \|a\| = \sup_{\psi \in \K^n, \psi \ne 0} \frac{\|a \psi\|}{\|\psi\|}  . \]
Every state on $O$ is of the form
\[     \omega(a) = \tr(a b)    \qquad \forall a \in O  \]
for some $b \in O$ with $b \ge 0$ and $\tr(b) = 1$.  We call such $b$ a \define{density matrix}.

Every skew-adjoint derivation of $O$ is of the form
\[       \lambda_a b = [a,b]  = ab - ba    \qquad \forall b \in O  \]
for some $a \in L$, and we have
\[       \exp(t \lambda_a) (b) = \exp(ta)\, b \, \exp(-ta) \qquad \forall t  \in \R, b \in O. \]
The transformations $\exp(t \psi_a)$ preserve the Jordan product and thus also the partial ordering on $O$ and unit $1 \in O$.   These transformations also act on any state $\omega$ to give a state $\omega \circ \exp(t \psi_a)$.    If $a$ generates time evolution, this is the usual time evolution of the state $\omega$, though in the complex case we usually write $a = iH$ for some $H \in O$, called the Hamiltonian.  
 
Every self-adjoint derivation of $O$ is of the form
\[        \delta_H b = H \circ b = \textstyle{\frac{1}{2}} (Hb + bH) \qquad \forall b \in O \]
for some $H \in O$, and we have 
\[    \exp(-\beta \delta_H) (b) = \exp(-\beta H/2) \, b \, \exp(-\beta H/2)  
 \qquad \forall \beta \in \R. \]
The transformations $\exp(-\beta \delta_H)$ preserve the partial ordering on $O$ but typically
not the Jordan product or the unit.   In Section \ref{sec:statistical} we argue that these transformations are connected to statistical mechanics.
\end{ex}

Given a unital JB-algebra $O$, Alfsen and Shultz \cite{AlfsenShultz98} found conditions under which its complexification
\[        \C \otimes O = \{a + ib: \; a,b \in O \}  \]
becomes a C*-algebra.   For this, they define a \define{dynamical correspondence} to be a linear map
\[           \psi \maps O \to L \]
obeying these conditions:
\begin{enumerate}[label=(\Alph*)]
\item    $\psi_a(a) = 0$ for all $a \in O$;
\item    $[\psi_a , \psi_b] = -[\delta_a,\delta_b] $ for all $a,b \in O$. 
\end{enumerate}
Condition (A) is a version of the self-conservation principle, since it
implies that $a$ is conserved by the one-parameter group generated by $\psi_a$:
\[        \exp(t \psi_a) (a) = a \qquad \forall t \in \R.  \]
It is equivalent to the statement
\[      \psi_a(b) = -\psi_b(a)    \qquad \forall a,b \in O  \]
which implies this version of Noether's theorem:
\[       \psi_a(b) = 0 \iff \psi_b(a)    \qquad \forall a,b \in O.\]
Condition (B) is a bit more mysterious, but let us state the theorem:

\begin{thm}[\textbf{Alfsen and Shultz}]
A unital JB-algebra $O$ is isomorphic to the self-adjoint part of a C*-algebra if and only if there exists a dynamical correspondence on $O$.   Each dynamical correspondence $\psi$ on $O$ determines a unique C*-algebra structure on $\C \otimes O$ with multiplication given by
\[    ab = a \circ b - i \psi_a(b)  \]
for all $a,b \in O$, $\ast$-structure such that elements of $O$ are self-adjoint and elements
of $L$ are skew-adjoint, and norm $\|a\| = \sqrt{\|a^*a\|}$ for all $a \in \C \otimes O$, where the norm in the square root is that in $O$.   Conversely, every C*-algebra structure on $\C \otimes O$ with these properties arises from a unique dynamical correspondence on $O$.
\end{thm}

\begin{proof}
This is Theorem 23 of Alfsen and Shultz \cite{AlfsenShultz98}; here we just mention some key 
steps.  The `only if' direction is easy since if $A$ is a C*-algebra and we define $O$ to be its Jordan algebra of self-adjoint elements and $L$ to be the Lie algebra of skew-adjoint derivations of $O$, the linear map
\[     
\psi \maps O \to L
\]
given by
\[      \psi_a(b) = \textstyle{\frac{i}{2}} [a,b]     \qquad \forall a ,b \in O \]
obeys conditions (A) and (B) in the definition of dynamical correspondence.  For the `if' direction, we take a dynamical correspondence $\psi$ and define an operation $\{-,-\} \maps O \times O \to O$ by 
\[    \{a,b\} = \psi_a(b)    \qquad \forall a,b \in O. \]
We use curly brackets here to emphasize that this is not a commutator. Then we use Theorem \ref{thm:complex_star-algebra} to make $\C \otimes O$ into a complex $\ast$-algebra.   Condition 1 of Theorem \ref{thm:complex_star-algebra} follows from condition (A).  Condition 2 follows from $O$ being a Jordan algebra.   Condition 3 follows from the fact that $\psi_a$ is a skew-adjoint derivation:
\[     \begin{array}{ccl}
    \{a, b \circ c\} &=& \psi_a(b \circ c) \\ 
&=&  \psi_a(b) \circ c + b \circ \psi_a(c) \\
&=& \{a, b\} \circ c + b \circ \{a,c\} . 
\end{array}
\]
Condition 4 follows from conditions (A) and (B):
\[   \begin{array}{ccl}
   (a \circ b) \circ c - a \circ (b \circ c) &=& c \circ (a \circ b) - a \circ (c \circ b) \\
   &=& [\delta_c,\delta_a] (b) \\
   &=& -[\psi_c, \psi_a](b) \\
   &=& -\{c,\{a,b\}\} + \{a,\{c,b\}\} \\
   &=&  \{\{a,b\},c\} - \{a,\{b,c\}\} 
\end{array}
\]
where (B) is used to show the antisymmetry of the bracket $\{-,-\}$.  To finish the job we need 
to check that $\C \otimes O$ is a C*-algebra, and that any C*-algebra structure on 
$\C \otimes O$ obeying the three properties listed in the theorem  arises from a unique 
dynamical correspondence.  The most technical part of Alfsen and Shultz's proof is checking the 
C*-axiom $\|a\|^2 = \|a^* a\|$ for all $a \in \C \otimes \O$.  \end{proof} 

In the finite-dimensional case Barnum and Hilgert proved a similar result for generalized probabilistic theories, where the convex set of states is taken as fundamental \cite[Prop.\ 9.2]{BarnumHilgert}; this result involves a map in the other direction, from generators to observables.  Earlier Barnum, Mueller and Ududec proved a similar result in terms of convex
cones \cite[Sec. VI]{BarnumMuellerUdudec}.   It would be interesting to extend such results to the infinite-dimensional case, especially because they do not assume a version of condition (B).  

\section{Statistical mechanics}
\label{sec:statistical}

Alfsen and Shultz showed that a unital JB-algebra $O$ comes from a complex C*-algebra if and only if it is equipped with a map from observables to generators, $\psi \maps O \to L$, obeying the self-conservation principle:
\[      \psi_a(a) = 0  \qquad \forall a \in O \]
and the mysterious condition (B):
\[           [\psi_a, \psi_b] = - [\delta_a, \delta_b]     \qquad \forall a,b \in O .\]
As we have seen, the self-conservation principle has a clear physical meaning connected to Noether's theorem.   We still need to understand condition (B).

When our JB-algebra comes from a C*-algebra we can take 
\[     \psi_a(b) = \textstyle{\frac{i}{2}}[a,b] \]
for all $a,b \in O$, and then condition (B) boils down to
\[          [ia, ib] = -[a,b]   \qquad  \forall a,b \in O. \]
So, there is no mystery from a purely mathematical viewpoint: condition (B) encodes the fact that $i^2 = -1$.  The question concerns the \emph{physical} meaning of this condition \emph{before} we take the complex numbers for granted.   Why should a map from observables to generators obey this condition?   

We do not have a completely satisfactory answer, but to make any progress on this question we need to better understand the self-adjoint order derivations $\delta_a$.   For a moment let us return to Example \ref{ex:matrix_algebras}, and assume that $O$ is the Jordan algebra of self-adjoint real, complex or quaternionic $n \times n$ matrices.  Choose $H \in O$ and treat it as the Hamiltonian of some system.   Then there is a self-adjoint order derivation $\delta_H$ on $O$ defined by
\[        \delta_H b = \textstyle{\frac{1}{2}}(Hb + bH)   \qquad \forall b \in O .\]
This generates a one-parameter group of transformations of $O$ such that
\[    \exp(-\beta \delta_H) (b) = \exp(-\beta H/2)\,  b \, \exp(-\beta H/2)   \qquad \forall \beta \in \R, b \in O .\]
If $\omega$ is a state on $O$, composing it with $\exp(-\beta \delta_H)$ typically does not give a state, but it gives a nonzero positive linear functional, which we can normalize to obtain a state $\omega_\beta$.  This is given by
\[           \omega_\beta(b) = \frac{1}{Z(\beta)}\, \omega(\exp(-\beta H/2)\,  b \, \exp(-\beta H/2)) \qquad \forall b \in O   \]
where 
\[         Z(\beta) =\omega(\exp(-\beta H)) . \]

There is a unique state invariant under all Jordan algebra automorphisms of $O$, given by
\[       \omega(b) = \frac{1}{n} \tr(b)   \qquad \forall b \in O .\]
In statistical mechanics its significance is that it maximizes the entropy.  In this case $Z(\beta)$ is called the \define{partition function}, and $\omega_\beta$ is the state of thermal equilibrium at temperature $T$ where $\beta = 1/kT$ and $k$ is Boltzmann's constant.  Explicitly,
\[        \omega_\beta(b) = \frac{\tr(e^{-\beta H} b)}{\tr(e^{-\beta H})}   \qquad \forall b \in O .\]
For each $\beta,\gamma \in \R$ we have
\[      \exp(-\beta \delta_H) \omega_\gamma = \frac{Z(\beta)}{Z(\gamma)} \omega_{\beta + \gamma}  .\]
Thus up to a factor involving the partition function, $\exp(-\beta \delta_H)$ acts as `translation in inverse temperature', somewhat as $\exp(-it \psi_H)$ acts as time translation.

The analogy between $it$ and $\beta = 1/kT$ is well known in physics.   It has even been enshrined as a principle: `inverse temperature is imaginary time'.    Under the name of `Wick rotation', this principle plays an important role in quantum field theory \cite{PeetersZamalkar}.  More speculatively, Hawking and many others have used it to study the thermodynamics of black holes \cite{FullingRuijsenaars}.  Condition (B) is a way of formulating this principle without explicitly mentioning the complex numbers.  As we have seen, the minus sign in this condition encodes the fact that $i^2 = -1$.  

Of course, the transformations $\exp(-\beta \delta_H)$ also act on states that are not in thermal equilibrium.  To understand how, it seems useful to look at Leifer and Spekkens' work on Bayesian inference in quantum theory  \cite{LeiferSpekkens}.   They define a product of nonnegative self-adjoint operators as follows:
\[         a \star b = a^{\frac{1}{2}} \, b \, a^{\frac{1}{2}}  \]
and they use this to describe the `Bayesian updating' of a state with density matrix $b$.   Given the state $\omega$ described by a density matrix $b$:
\[    \omega(a) = \tr(ab)  \qquad \forall a \in O \]
we have
\[    \omega_\beta(a) = \frac{1}{Z(\beta)} \tr(a (\exp(-\beta H) \star b))  \qquad \forall a \in O.\]
We can roughly say that $\omega_\beta$ is the result of updating $\omega$ by multiplying the probability of being in any state of energy $E$ by the Boltzmann factor $\exp(-\beta E)$.

Returning from the example to the case of an arbitrary unital JB-algebra, it is worth noting that
the $\star$ product is a special case of McCrimmon's \cite{McCrimmon} operation 
\[    U_a(b) = 2(a \circ (a \circ b)) - (a \circ a) \circ b   \qquad \forall a,b \in O \]
since we can write
\[         a \star b = U_{a^{\frac{1}{2}}} (b)   \]  
where the square root is well-defined using the functional calculus whenever $a \ge 0$.   This formula for the $\star$ product reduces to Leifer and Spekkens' definition when $O$ is a special Jordan algebra.  If we generalize their ideas to arbitrary unital JB-algebras in this way, we may get a better understanding of the physical meaning of self-adjoint order derivations.

To conclude, it is worth noting that there is a link between self-adjoint and skew-adjoint
order derivations even before we have a dynamical correspondence.  Recall that there is a Lie algebra of \emph{all} order derivations
\[             A = O \oplus L   \]
with $O$ consisting of the self-adjoint order derivations and $L$ consisting of skew-adjoint ones.
It is easy to check that the bracket in $A$ has
\[    [-,-] \maps L \times L \to L    \qquad [-,-] \maps L \times O \to O \]
\[    [-,-] \maps O \times L \to O    \qquad [-,-] \maps O \times O \to L .\]
Mathematically, we summarize this by saying that $A$ is a $\mathbb{Z}/2$-graded Lie 
algebra with $L$ as its `even' part and $O$ as its `odd' part.   In particular, the commutator of 
two self-adjoint order derivations $\delta_a$ and $\delta_b$ is skew-adjoint.  This is what makes it \emph{possible} to demand condition (B).

Since groups are conceptually simpler than Lie algebras, it is good to note that $A$ is the Lie algebra of $\Aut_{\ge}(O)$, the Banach--Lie group of order automorphisms of $O$, while $L$ is the Lie algebra of $\Aut(O)$, the Banach--Lie group of Jordan algebra automorphisms of $O$.  The smaller group is connected to quantum mechanics, while the larger group is connected to both quantum and statistical mechanics.   In the finite-dimensional case, the fact that $A$ is a $\mathbb{Z}/2$-graded Lie algebra makes the quotient space $\Aut_{\ge}(O)/\Aut(O)$ into a `symmetric space'.  This space is just the interior of the convex cone $O^+$ \cite{FarautKoranyi}.  Something similar holds in the infinite-dimensional case, though this is somewhat less studied \cite{Chu2017}.

Thus, in any unital JB-algebra, even before imposing extra conditions, there is a deep interplay between quantum and statistical mechanics.   It remains to more deeply understand this interplay and its connection to Noether's theorem and the role of the complex numbers in quantum mechanics.

\end{document}